\documentclass[hyper]{JHEP} % 10pt is ignored!

\usepackage{epsfig}
\newcommand\fverb{\setbox\pippobox=\hbox\bgroup\verb}

\newcommand\fverbdo{\egroup\medskip\noindent%

            \fbox{\unhbox\pippobox}\ }

\newcommand\fverbit{\egroup\item[\fbox{\unhbox\pippobox}]}

\newbox\pippobox

\title{
Ho\v{r}ava-Lifshitz $f(R)$ Gravity}
\author{J. Kluso\v{n}\\
Department of
Theoretical Physics and Astrophysics\\
Faculty of Science, Masaryk University\\
Kotl\'{a}\v{r}sk\'{a} 2, 611 37, Brno\\
Czech Republic\\
E-mail: \email{klu@physics.muni.cz}}
\preprint{\hepth{0907.3566}}

 \abstract{This paper is devoted to the construction
of new type of $f(R)$ theories of gravity that
are  based on the principle of detailed balance. We
discuss two versions of these theories with and without
the projectability condition.}

\def\hc{\hat{c}}
\def\mH{\mathcal{H}}

\def\iD{\left(D^{-1}\right)}

\def\bx{\mathbf{x}}
\def\by{\mathbf{y}}

\newcommand{\hQ}{\hat{Q}}
\newcommand{\hPhi}{\hat{\Phi}}
\newcommand{\hPi}{\hat{\Pi}}

\newcommand{\mG}{\mathcal{G}}

\def\ket #1{\left|#1\right>}

\newcommand{\bT}{\mathbf{T}}

\newcommand{\mL}{\mathcal{L}}

\def\pb #1{\left\{#1\right\}}

\begin{document}
%%%%%%%%%%%%%%%%%%%%%
%%%%Introduction %%%%%%%%%
%%%%%%%%%%%%%%%%%%%%
\section{Introduction}\label{first}
Recently in series of very interesting
papers P. Ho\v{r}ava suggested new
approach for the study of membranes and
quantum gravity theories known as
Ho\v{r}ava-Lifshitz gravities
\cite{Horava:2009if,Horava:2009uw,Horava:2008ih,Horava:2008jf}
\footnote{ Ho\v{r}ava's ideas were
elaborated from different points of
view in couple of papers, see for
example
\cite{Appignani:2009dy,Adam:2009gq,Wang:2009yz,Bogdanos:2009uj,Kobakhidze:2009zr,
Peng:2009uh,Mukohyama:2009tp,Castillo:2009ci,Iengo:2009ix,
Cai:2009hc,Das:2009ba,Blas:2009yd,Cai:2009in,Shu:2009gc,Ghodsi:2009zi,Germani:2009yt,
Myung:2009sa,BottaCantcheff:2009mp,Park:2009zr,Sakamoto:2009ww,Nojiri:2009th,Kocharyan:2009te,
Wang:2009rw,Calcagni:2009qw,Mukohyama:2009mz,Minamitsuji:2009ii,Gao:2009ht,Saridakis:2009bv,
Kim:2009zn,Bertoldi:2009vn,Dhar:2009dx,Sotiriou:2009bx,Li:2009bg,Pang:2009ad,Charmousis:2009tc,
Chen:2009jr,Konoplya:2009ig,Kluson:2009hr,Chen:2009gs,Orlando:2009en,Cai:2009dx,Sotiriou:2009gy,
Gao:2009bx,Cai:2009ar,Cai:2009pe,Nastase:2009nk,Brandenberger:2009yt,Mukohyama:2009gg,
Lu:2009em,Kluson:2009sm,Kiritsis:2009sh,Calcagni:2009ar,Visser:2009fg,
Colgain:2009fe,Chen:2009bu,Cai:2009qs,Takahashi:2009wc,Piao:2009ax}.}.
The attractive property of
Ho\v{r}ava-Lifshitz gravity is that it
is power-counting renormalizable. The
second important property of the
Ho\v{r}ava construction is
\emph{detailed balance condition}. This
condition is based on the idea that the
potential term in the Lagrangian of
$D+1$ dimensional theory descants from
the variation of $D$ dimensional
action. In fact, this construction is
based on the following idea known from
the condensed matter physics
\cite{Ardonne:2003wa}: \emph{Is it
possible to find such a $D+1$
dimensional quantum theory such that
its ground state wave functional
reproduces the partition function of
$D$ dimensional theory?} This idea was
elaborated in details in
\cite{Ardonne:2003wa} and recently in
series of papers by P. Ho\v{r}ava with
many interesting results. In
particular, if we start with known
classical universality classes in $D$
dimension we can construct a quantum
critical systems in $D+1$ dimensions.

It is very interesting that similar
situation naturally occurs in case of
topological string theory
\cite{Witten:1993ed,Gunaydin:2006bz},
OSV conjecture, topological M-theory,
together with non-critical M-theory
\cite{Dijkgraaf:2004te,Horava:2005tt,Horava:2005wm}
\footnote{For recent discussion and
extensive list of references, see
\cite{Dijkgraaf:2009gr}.}.

As was carefully discussed recently in
\cite{Mukohyama:2009tp}, there  are at
least four versions of the theory:
with/without the detailed balance
condition; and with/without the
 projectability condition. As we will
 show bellow the projectability
 condition means that the lapse
 function depends on time only.
 It was argued in many papers that
the most promising is the version
without the detailed balance condition
with the projectability condition that
has  a potential to be theoretically
consistent and cosmologically viable.

Even if there are doubts considering
the detailed balance condition in
general relativity we feel that it
deserves to be studied further. In
particular, let us consider following
situation when we have a $D$ dimensional
quantum  theory with
 corresponding partition function
$\mathcal{Z}$.
Then we consider Hamiltonian of
$D+1$ dimensional theory   and ask
the question under which condition this
Hamiltonian
annihilates the vacuum wave functional
of given $D+1$ dimensional theory
 where the norm of this vacuum state
coincides with the partition function
$\mathcal{Z}$. We show that there
exists an infinite number of such
Hamiltonians that can be defined as a
Taylor series in powers of creation and
annihilation operators where the
annihilation operator annihilates the
vacuum wave functional.

We apply this  construction to the case of
theory of gravity and consider two
situations. In the first one we follow the
original Ho\v{r}ava's approach
\cite{Horava:2008ih} where we start
with partition function of
 $D$ dimensional gravity and
 demand that there exists quantum gravity
 in $D+1$ dimension such that the
 norm of the ground state functional
 coincides with the partition function
 of $D$ dimensional theory. Then
 we show that we can construct infinite
 number of Hamiltonians that annihilate
 this ground state. In other words
 we find an infinite number of Hamiltonians
that obey the detailed balance
conditions. Clearly these Hamiltonians
are well defined at the classical level
due to the well known peculiarities
that arise in non-linear quantum
theories. Even at the classical level
these Hamiltonians have many
interesting properties that should be
studied further. For example, for the
special form of the Hamiltonian that
will be specified below we determine
 corresponding
Lagrangian and we find that the  action
for this theory  is manifestly
invariant under spatial diffeomorphism.
Then, following \cite{Horava:2008ih} we
perform an extension of given
symmetries that leads to the action
that is invariant under \emph{foliation
preserving diffeomorphism}. We find new
non-linear theory of gravity that
resemble $f(R)$ theory of gravity
\footnote{For review and extensive list
of references, see
\cite{Faraoni:2008mf,
Nojiri:2006ri,Nojiri:2008nt}.} that
however is not invariant under full
$D+1$ diffeomorphism. It would be
certainly interesting  to study the
cosmological implications of this model
exactly in the same way as in case of
ordinary $f(R)$ theory of gravity.
Certainly we can also consider the more
general form of Hamiltonians then the
Hamiltonian explicitly studied in this
paper.

In the second case we consider an
alternative form of the principle of
detailed balance. We  consider the
 situation when the Hamiltonian
density of $D+1$ dimensional theory is
a linear combination of the
diffeomorphism and Hamiltonian
constraint. Further we assume that the
 Hamiltonian constraint has the
special property that  it annihilates
the vacuum wave functional that has the
norm equal to the partition function of
$D$ dimensional theory of gravity.
Since this vacuum wave functional is
manifestly invariant under $D$
dimensional
 spatial diffeomorphisms
it is  annihilated by the generator of
diffeomorphism and consequently by the
Hamiltonian of the theory. We would
like to stress that at this moment we
only assume that the Hamiltonian
annihilates the vacuum wave functional
but we do not demand that it should
annihilate all states of the theory. On
the other hand we will argue that the
Hamiltonian framework implies that the
Hamiltonian should annihilate all
states in case of the quantum
mechanical formulation of the theory.
Explicitly, following the standard
approach we determine an
 action corresponding to given
 Hamiltonian.  Then we
continue in the study of this theory
and develop the Hamiltonian formalism
that follows from this action. Since
the action contains the fields
$N(t,\bx)$ and $N_i(t,\bx)$ without
time derivatives we find that the
absence of corresponding momenta imply
the primary constraints of the theory
$\pi_N(\bx)\approx 0 \ ,
\pi^i(\bx)\approx 0$.  The consistency
of these constraints with the time
evolution of the system implies the
secondary constraints
$\mH_0(\bx)\approx 0 \ , \quad
\mH^i(\bx)\approx 0$. Following the
Dirac approach we then find that
\emph{all} quantum states of the theory
have to be annihilated by these
constraints as opposite to the original
assumption that the state that should
be annihilated by $H$ is the vacuum
state only. On the other hand we will
argue that
 these
theories suffer from the same problems
as the Ho\v{r}ava's theories without
projectability conditions
\cite{Li:2009bg}. However using the
fact  that these theories are
constructed as theories that obey the
detailed balance conditions it is
possible to  find the algebra of
constraints that close however that do
not support any physical excitations at
all. In other words these theories are
topological. As the second example of
solvable theory we consider the case of
ultralocal theory and we argue  the
algebra of constraints closes  as in
standard gravity theory.

Let us outline our results. Imposing
the detailed balance condition we are
able to find new $D+1$  $f(R)$ theories
of gravity with or without
projectability conditions. We would
like to stress that these theories
should be considered as toy models of
gravity theories. It would be
interesting to study the cosmological
implications of these theories in the
same way  as in case of $f(R)$ theories
of gravity with full diffeomorphism
invariance.

This paper is organized as follows. In
next section (\ref{second}) we present
the main idea of our construction on
the simple case of collection of $D$
scalar fields in $p+1$ dimensions. Then
in section (\ref{third}) we perform the
construction of new $D+1$ dimensional
theory of gravity that is invariant
under foliation preserving
diffeomorphism. In section
(\ref{fourth}) we suggest an
alternative way how to impose the
condition of the detailed balance in
case of $D+1$ dimensional theory of
gravity and we argue that this
procedure leads to $D+1$ dimensional
theory without projectability
condition.

%%%%%%%%%%%%%%%%%%%%%%%%%%%%%%%%%%%%%%%%%%%%
\section{Non-Linear Scalar Lifshitz Theory }
\label{second} In this section we
describe  the
construction of non-linear Lifshitz theory
based on the detailed balance condition
on the simple example of collection of
$D$ scalar fields in $p$ dimensions.
 This procedure is based on an idea is
that the norm of the  ground state
functional of $p+1$ dimensional  theory
coincides with the  partition function
of any $p$ dimensional theory. We
should  stress that   this requirement
is pure formal since  we do not carry
about issues whether this partition
function is well defined.  Very
nice discussion of  issues that are
related to the construction of wave
functionals  can be found in paper
\cite{Witten:2003mb}. Despite of this
fact we proceed further and we find
that we are able to find new
interesting class of theories at least
at the classical level.

Let us start in the same way as in
\cite{Horava:2008ih,Ardonne:2003wa} and
consider  the situation when we have a
collection of $D$ scalar fields defined
on $p$ dimensional Euclidean space with
coordinates $\bx=(x^i) \ , i=1,\dots,p$
with following  action
\begin{equation}\label{Waction}
W=\frac{1}{2}\int d^p\bx \delta^{ij}
\partial_i\Phi^M\partial_j \Phi^N
g_{MN} \ ,
\end{equation}
where  $g_{MN}\ , M,N=1,\dots,D$ is a
\emph{constant positive definite
symmetric} matrix. Clearly we can
consider more general form of the
action than the one given in
(\ref{Waction}) but in order to explain
the main idea of the constructions we
restrict ourselves to the simple action
 given above.

As in standard quantum mechanics the
fundamental object of this
 theory is the
partition function $\mathcal{Z}$
\begin{equation}\label{Zi}
\mathcal{Z}=\int \mathcal{D} \Phi(\bx)
\exp[-W(\Phi(\bx))]
\end{equation}
that is defined as
 a path integral on the space of
field configurations $\Phi^M(\bx)$.
Then let us  assume an existence of
$p+1$ dimensional quantum field theory
with collection of the operators
$\hat{\Phi}^M(\bx)$ and their conjugate
momenta $\hat{\Pi}_M(\bx)$ and that
obey the canonical commutation relation
\begin{equation}\label{com}
[\hat{\Phi}^M(\bx),\hat{\Pi}_N(\by)]=
i\delta^M_N\delta(\bx-\by) \ .
\end{equation}
Further, we introduce  eigenstate of
$\hat{\Phi}^M(\bx)$ that is  the state
$\ket{\Phi(\bx)}$ that obeys
\begin{equation}
\hat{\Phi}^M(\bx)\ket{\Phi(\bx)}=
\Phi^M(\bx)\ket{\Phi(\bx)} \ .
\end{equation}
In the Schr\"{o}dinger  representation
any state of given system is
represented as the state functional
$\Psi[\Phi(\bx)]$ and  the standard
interpretation of quantum mechanics
implies that $\Psi[\Phi(\bx)]
\Psi^*[\Phi(\bx)]$ is a density on the
configuration space. Note also that
action of the operator
$\hat{\Phi}^M(\bx)$ on this state
functional corresponds to
multiplication with $\Phi^M(\bx)$. On
the other hand the commutation relation
(\ref{com}) implies that in the
Schr\"{o}dinger representation the
operator $\hat{\Pi}_M(\bx)$ is equal to
\begin{equation}
\hat{\Pi}_M(\bx)=-i\frac{\delta}{\delta
\Phi^M(\bx)} \ .
\end{equation}
 Our goal is to
formulate $p+1$ dimensional system with
the property that the norm of its
ground-state functional
$\Psi_0[\Phi(\bx)]$ reproduces the
partition function (\ref{Zi})
\begin{equation}
\left<\Psi_0|\Psi_0\right>= \int
\mathcal{D}\Phi(\bx)
\Psi_0^*[\Phi(\bx)] \Psi_0[\Phi(\bx)]
=\int \mathcal{D} \Phi(\bx)
\exp[-W(\Phi(\bx))] \ .
\end{equation}

Everything that has been done up to
this point is well known. However we
now make a presumption that the
Hamiltonian of $p+1$ dimensional theory
has the form
\begin{equation}\label{defH1}
\hat{\mH}(\bx)=\kappa^2\left(
\sum_{n=0}^\infty \hc_n(\hPhi)
(\hQ^\dag_M g^{MN}
\hQ_N)^n-\hc_0(\hPhi)\right) \ ,
\nonumber \\
\end{equation}
where $\kappa$ is a coupling constant,
$\hc_n(\hPhi)$ are functions that
generally depend on the operators
$\hPhi$ and
 where $\hat{Q}_M,\hat{Q}_M^\dag$ are
defined as
\begin{equation}\label{defQ}
\hat{Q}_M=i\hat{\Pi}_M+\frac{1}{2}\frac{\delta
W[\hat{\Phi}]}{\delta
\hat{\Phi}^M(\bx)} \ , \quad
\hat{Q}_M^\dag=
-i\hat{\Pi}_M+\frac{1}{2}\frac{\delta
W[\hat{\Phi}]}{\delta
\hat{\Phi}^M(\bx)} \ .
\end{equation}
%Clearly, the Hamiltonian (\ref{defH})
%is  Hermitian and positive.
 Note that in the Schr\"{o}dinger
 representation the operators
 $\hat{Q}_M,\hat{Q}_M^\dag$ are equal
 to
\begin{equation}\label{QMshch}
\hat{Q}_M=\frac{\delta}{\delta\Phi^M(\bx)}+\frac{1}{2}\frac{\delta
W[\Phi]}{\delta \Phi^M(\bx)} \ , \quad
\hat{Q}_M^\dag=
-\frac{\delta}{\delta\Phi^M(\bx)}+\frac{1}{2}\frac{\delta
W[\Phi]}{\delta \Phi^M(\bx)} \ .
\end{equation}
Let us assume that   the vacuum wave
functional takes the form
\begin{equation}
\Psi_0[\Phi(\bx)]=
\exp\left(-\frac{1}{2}W\right)
=\exp\left(-\frac{1}{4}\int d^p\bx
\delta^{ij}\partial_i\Phi^M(\bx)g_{MN}\partial_j
\Phi^N(\bx)\right) \ .
\end{equation}
Then  it is easy that $\hat{Q}_M$
defined in (\ref{QMshch}) annihilates
$\Psi_0$
\begin{equation}
\hat{Q}_M\Psi[\Phi(\bx)]=0
\end{equation}
as follows from the fact that
\begin{equation}
i\hat{\Pi}(\bx)
\Psi_0[\Phi]=
\frac{\delta }{\delta \Phi^M(\bx)}
\Psi_0[\Phi]=-\frac{1}{2}\frac{\delta
W}{\delta \Phi^M(\bx)}\Psi_0[\Phi] \ .
\end{equation}
In other words the vacuum wave
functional is annihilated by $\hQ_M$
and by construction  it is an
eigenstate of the Hamiltonian with zero
energy. Further, the norm of the vacuum
wave functional coincides with the
partition function of $p$ dimensional
theory.
% strictly keep the ordering as suggested
% above. Explicitly, we define the
% ordering of the operators
% $\hat{c}_n(\hPhi)$ and
% $\hQ_M,\hQ_M^\dag$  as
% \begin{equation}
% \hat{c}_1 (\hQ^{\dag M}\hQ_M) \ , \quad
% \hat{c}_2(\hQ^{\dag M}\hQ_M)(\hQ^{\dag
% N}\hQ_N) \ , \dots \ .
% \end{equation}
%\begin{eqnarray}
%H\Psi=\int d^D\bx
%\mH\Psi=c_0 \int d^D\bx \Psi
%\nonumber \\
%\end{eqnarray}

It is clear that it this way we can
define an infinite number of
Hamiltonians that obey the detailed
balance condition. In what follows we
restrict ourselves to the  following
 example of the Hamiltonian
density
\begin{equation}\label{Hdef}
\hat{\mH}=\kappa^2\left(
\sqrt{\hat{\alpha}(\hPhi)+\hat{
\beta}(\hPhi)\left(\hPi_Mg^{MN}\hPi_N
+\frac{1}{4}\left(\frac{\delta
W}{\delta\hPhi^M}g_{MN}\frac{\delta
W}{\delta\hPhi^N}\right)\right)}-\sqrt{\hat{\alpha}(\hPhi)}\right)
\ ,
\end{equation}
where $\hat{\alpha},\hat{\beta}$
generally depend on $\hPhi$ and where
the square root function in the
definition of the Hamiltonian is
defined as the Taylor polynomial in
$(\hQ^\dag \hQ)^n$ written explicitly
in  (\ref{defH1}).

As the next step we determine the
Lagrangian from the classical form
of the  Hamiltonian density
(\ref{Hdef}). Using the Hamiltonian
equation $\partial_t \Phi=\pb{\Phi,H}$
and the form of the Hamiltonian density
(\ref{Hdef}) we find
\begin{eqnarray}
\partial_t \Phi^M=\pb{\Phi^M,H}=
\kappa^2\frac{\beta\Pi_Ng^{NM}}{\sqrt{\alpha+\beta(\Pi_Mg^{MN}
\Pi_N+ \frac{1}{4}\frac{\delta
W}{\delta\Phi^M}g^{MN}
\frac{\delta W}{\delta\Phi^N})}} \   \nonumber \\
% \Pi^Mg_{MN}\Pi^N=(\alpha+\beta\frac{1}{4}\frac{\delta
% W}{\delta\Phi^M}g^{MN}\frac{\delta
% W}{\delta\Phi^N})
% \frac{\partial_t\Phi^Mg_{MN}\partial_t\Phi^N}{(
% \kappa^4\beta^2-\beta\partial_t\Phi^Mg_{MN}\partial_t\Phi^N)}
% \ . \nonumber \\
\end{eqnarray}
so that  the Lagrangian
density is equal to
\begin{eqnarray}\label{Lifshitzgen}
\mL&=&\Pi_M\partial_t
\Phi^M-\mH=\nonumber \\
% =-(\alpha+\frac{\beta}{4}
% \frac{\delta
% W}{\delta\Phi^M}g^{MN}\frac{\delta
% W}{\delta\Phi^N})\frac{1}{\sqrt{\alpha+\beta(
% \Pi_Mg^{MN}\Pi_N
% +\frac{1}{4}\frac{\delta
% W}{\delta\Phi^M}g_{MN}\frac{\delta
% W}{\delta\Phi^N})}}+\kappa^2\sqrt{\alpha}= \nonumber
% \\
&=&-\kappa^2\sqrt{\alpha(\Phi)
+\frac{\beta(\Phi)}{4}\frac{\delta
W}{\delta\Phi^M}g^{MN} \frac{\delta
W}{\delta\Phi^N}}
\sqrt{1-\frac{1}{\kappa^4\beta(\Phi)}
\partial_t\Phi^M
g_{MN}\partial_t\Phi^N}+\kappa^2\sqrt{\alpha(\Phi)} \ .
 \nonumber
\\
\end{eqnarray}
Let us now simplify the action further
and consider the case when
 $\alpha=1,\beta=\mathrm{const}$. Then,
since the variation of (\ref{Waction})
is
 equal to
\begin{equation}
\frac{\delta W}{\delta\Phi^M(\bx)}=
-\partial^i\partial_i\Phi^N(\bx)g_{NM}
\end{equation}
we find that the  action of $p+1$
dimensional theory takes the form
\begin{eqnarray}
S=-\kappa^2\int d^p\bx dt
\sqrt{1+\frac{\beta}{4}
(\partial^i\partial_i\Phi^M)
g_{MN}(\partial^j\partial_j\Phi^N)}
\sqrt{1-\frac{1}{\kappa^4\beta}
\partial_t\Phi^Mg_{MN}
\partial_t\Phi^N} \ .  \nonumber \\
\end{eqnarray}
If we now expand this action up to
quadratic order in fields we find the
standard Lifshitz action (up to trivial
rescaling of $\beta$)
\begin{equation}
S=-\kappa^2
%\sqrt{\alpha}
\int dt d^p\bx -\int dt d^p\bx[
%\frac{
\kappa^2\beta
%{\sqrt{\alpha}}
\frac{1}{8}
(\partial^i\partial_i\Phi^M)g_{MN}
(\partial^j\partial_j \Phi^N)
-
%\frac{\sqrt{\alpha}
\frac{1}{ 2\kappa^2\beta}
\partial_t
\Phi^Mg_{MN}\partial_t\Phi^N] \ .
\end{equation}
In other words for small spatial and
time derivatives the Lagrangian
(\ref{Lifshitzgen}) reduces to the
Lifshitz scalar theory.
%%%%%%%%%%%%%%%%%%%%%%%%%%%%%%%%%%%%%%%%%
\section{Ho\v{r}ava-Lifshitz $f(R)$ Theory of
Gravity-With Projectability
Condition}\label{third} Let us now turn
to the main topic of this paper which
is a  construction of the
Ho\v{r}ava-Lifshitz $f(R)$ theories of
gravity in $D+1$ dimensions. This
construction is based on assumption
that we have $D+1$ dimensional quantum
 theory of gravity that is characterized
 by following quantum
Hamiltonian density
\begin{eqnarray}\label{QGRH}
\hat{\mH}=\kappa^2 \sqrt{\hat{g}}
\left(\sum_{n=0}^\infty \hc_n(\hat{g}_{ij})
(\hat{Q}^{\dag ij} \frac{1}{\hat{g}}
\hat{\mG}_{ijkl}\hat{Q}^{kl})^n-
\hc_0(\hat{g}_{ij})\right) \ ,  \nonumber \\
\end{eqnarray}
where
\begin{equation}\label{hQij}
\hat{Q}^{\dag ij}= -i\hat{\pi}^{ij}
+\sqrt{\hat{g}}\hat{E}^{ij}(\hat{g}_{ij})
\ , \quad \hat{Q}^{\dag
ij}=-i\hat{\pi}^{ij}
+\sqrt{\hat{g}}\hat{E}^{ij}(\hat{g}_{ij})
\ ,
\end{equation}
and where $\hat{g}=\det \hat{g}_{ij}$
and $\kappa$ is a coupling constant of
given theory. Note that the fundamental
operators of quantum theory of gravity
are  metric components
 $\hat{g}_{ij}(\bx) \ ,
i=1,\dots,D$ together with their conjugate
momenta $\hat{\pi}^{ij}(\bx)$. These
operators  obey
the commutation relations
\begin{equation}
[\hat{g}_{ij}(\bx),\hat{\pi}^{kl}(\by)]=
\frac{1}{2}(\delta_i^k\delta_j^l+\delta_i^l\delta_j^k)
\delta(\bx-\by) \ .
\end{equation}
Further, $\hat{c}_n$ defined in
(\ref{QGRH}) are scalar functions that
depend on
$\hat{g}_{ij}$ only.
 It is
clear that in the Schr\"{o}dinger
representation the operators
(\ref{hQij}) take the form
\begin{equation}
\hat{Q}^{ij}(\bx)=-\frac{\delta}
{\delta g^{ij}(\bx)}+\sqrt{g}(\bx)E^{ij}(\bx) \ ,
\quad
\hat{Q}^{\dag ij}(\bx)=
\frac{\delta }{\delta  g^{ij}(\bx)}+
\sqrt{g}(\bx)E^{ij}(\bx) \ .
\end{equation}
The next goal is to specify the form of
the operators $E^{ij}$. To do this we
assume that the theory obeys the
\emph{detailed balance condition} so
that
\begin{equation}\label{defE}
\sqrt{g}(\bx)E^{ij}(\bx)=\frac{1}{2}\frac{
\delta W}{\delta g^{ij}(\bx)} \ ,
\end{equation}
where $W$ is an action of $D$
dimensional gravity. As in
\cite{Horava:2008ih}
 we construct the vacuum wave functional of
$D+1$ dimensional theory as
\begin{equation}\label{vvf}
\Psi[g(\bx)]=
\exp\left(-\frac{1}{2}W\right) \ ,
\end{equation}
where $W$ is the Einstein-Hilbert
action in $D$ dimensions
\begin{equation}
W=\frac{1}{2\kappa^2_W}
\int d^D \bx \sqrt{g}R \ .
\end{equation}
Generally the action $W$ could also
contains additional terms that are
functions of metric however the
explicit form of $W$ will not be
important in following discussion.

The form of the vacuum wave functional
(\ref{vvf}) implies that it is
annihilated by
 (\ref{QGRH}). Further as a consequence
of the detailed balance condition  the norm of the
 functional (\ref{vvf})  coincides with the partition
 function of $D$ dimensional Euclidean gravity.
 In other words we have again infinite
 number of possible Hamiltonians that
annihilate the vacuum state (\ref{vvf})
and that are defined using the
principle of detailed balance.

In order to find the Lagrangian
formulation of this  theory we now
consider the classical form of the
Hamiltonian density (\ref{QGRH}).
In order to simplify the analysis
 we restrict ourselves to the
following explicit form  of the Hamiltonian
density
\begin{eqnarray}\label{Hdecl}
\mH=\kappa^2
\sqrt{g}\left(\sqrt{1+
\beta
(-i\pi^{ij}+
\sqrt{g}E^{ij})\frac{1}{g}
\mG_{ijkl}
(i \pi^{kl}+ \sqrt{g}E^{kl})}-1\right) \ , \nonumber \\
\end{eqnarray}
where $\mG_{ijkl}$ denotes the inverse
of the De Witt metric
\begin{equation}\label{DeWi}
\mG_{ijkl}=\frac{1}{2}(g_{ik}g_{jl}+
g_{il}g_{jk})-\tilde{\lambda}g_{ij}g_{kl}
\
\end{equation}
with
$\tilde{\lambda}=\frac{\lambda}{D\lambda-1}$.
The "metric on the space of
metric", $\mathcal{G}^{ijkl}$ is
defined as
\begin{equation}\label{DeW}
\mG^{ijkl}=\frac{1}{2}(g^{ik}g^{jl}+g^{il}
g^{jk}-\lambda g^{ij}g^{kl}) \
\end{equation}
with $\lambda$ an arbitrary real
constant. Note that (\ref{DeWi})
together with (\ref{DeW}) obey the
relation
\footnote{
Note that we use the terminology
introduced in \cite{Horava:2008ih}
and that we review there.
 In case of relativistic
theory, the full diffeomorphism
invariance fixes the value of $\lambda$
uniquely to equal $\lambda=1$. In this
case the object $\mG_{ijkl}$ is known
as the "De Witt metric". We use this
terminology to more general case when
$\lambda$ is not necessarily equal to
one.}
\begin{equation}
\mG_{ijmn}\mG^{mnkl}=\frac{1}{2}
(\delta_i^k\delta_j^l+\delta_i^l\delta_j^k)
\ .
\end{equation}
The form  of the Hamiltonian
density (\ref{Hdecl}) implies
following
time derivative of $g_{ij}$
\begin{eqnarray}
\partial_t g_{ij}=\pb{g_{ij},H}=
\kappa^2 \frac{\beta\mG_{ijkl}\pi^{kl}}
{\sqrt{g}\sqrt{1+ \beta (-i
\pi^{ij}+\sqrt{g}
E^{ij})\frac{1}{g}\mG_{ijkl}
(i\pi^{kl}+ \sqrt{g}E^{kl})}} \ . \nonumber \\
\end{eqnarray}
With the help of this result
we can express  $\pi^{ij}$ as
a function of $g_{ij}$ and $\partial_t
g_{ij}$. Then we  easily find  the
corresponding  Lagrangian density
in the form
\begin{eqnarray}\label{Ld}
\mL=\partial_t g_{ij}\pi^{ij}-\mH=
-\kappa^2\sqrt{g}\left(\sqrt{1+\beta
E^{ij}\mG_{ijkl} E^{kl}}
\sqrt{1-\frac{1}{\kappa^4\beta}
\partial_t g_{ij}\mG^{ijkl}
\partial_t g_{kl}}-1\right) \ .  \nonumber
\\
\end{eqnarray}
By construction the action
\begin{equation}\label{SHGG}
S=\int d^D \bx \mL \ ,
\end{equation}
 where  $\mL$ is  given in
(\ref{Ld}) is invariant under the
global time translation $t'=t+\delta t
\ , \delta t=\mathrm{const}$ and under
the spatial diffeomorphism
\begin{equation}\label{spd}
x'^i=x^i(\bx) \ .
\end{equation}
This follows from the fact that we
presumed that
 the functional $W$ is
invariant under the spatial
diffeomorphism under which the metric
$g_{ij}$
 and tensor $E^{ij}$ transform as
\begin{eqnarray}\label{mettr}
g'_{ij}(\bx')&=&g_{kl}(\bx)\iD^k_i
\iD^l_j  \ , \nonumber \\
E'^{ij}(\bx')&=&
E^{kl}(\bx)D_k^i D_l^j \ ,\nonumber \\
\end{eqnarray}
where
\begin{equation}
 D^i_j=\frac{\partial
x'^i}{\partial x^j} \ , \quad D^i_j
\iD^j_k=\delta^i_k \ .
\end{equation}
%\begin{eqnarray}
%\frac{\delta W'}{\delta g'_{ij}(\bx')}
%=\frac{\delta W}{\delta g_{kl}(\bx)}
%\frac{\delta g_{kl}(\bx)} {\delta
%g'_{ij}(\bx')}= \frac{\delta W}{\delta
%g_{kl}(\bx)}D^k_i D^l_j
% \nonumber \\
%\end{eqnarray}
Using the transformation property of
$g_{ij}$ we find that the metric
 $\mG_{ijkl}$ transforms
as
\begin{eqnarray}
\mG'_{ijkl}(\bx')=
% \frac{1}{2}(g_{i'k'}\iD^{i'}_i
% \iD^{k'}_k g_{j'l'}\iD^{j'}_j
% \iD^{l'}_l+\nonumber \\
% g_{i'l'}\iD^{i'}_i \iD^{l'}_l
% g_{j'k'}\iD^{j'}_j
% \iD^{k'}_k)-\tilde{\lambda}g_{i'j'}
% \iD^{i'}_i\iD^{j'}_j
%  g_{k'l'}\iD^{k'}_k\iD^{l'}_l=
% =\nonumber \\
\mG_{i'j'k'l'}(\bx)\iD^{i'}_i
\iD^{j'}_j \iD^{k'}_k\iD^{l'}_l \
\nonumber \\
\end{eqnarray}
and the invariance of the action under
the  spatial diffeomorphism (\ref{spd})
is obvious.
%These results imply  that $E^{ij}
%\mG_{ijkl} E^{kl}$ and  $\partial_t
%g_{ij}\mG^{ijkl}\partial_tg_{kl}$ are
%invariant under spatial diffeomorphism.
%In summary, we find that the action
%\begin{equation}\label{SHGG}
%S=-\kappa^2\int d^D\bx\sqrt{g}
%\left(\sqrt{1+\beta E^{ij}\mG_{ijkl}
%E^{kl}} \sqrt{1-\frac{1}{\kappa^4\beta}
%\partial_t g_{ij}\mG^{ijkl}
%\partial_t g_{kl}}-1\right)
%\end{equation}
% is invariant under spatial
%diffeomorphism.
%%%%%%%%%%%%%%%%%%%%%%%%%%%%%%%%%%%%%%%
\subsection{Extension of Symmetries}
We  argued that the action formulated
above is invariant under
 $D$ dimensional \emph{spatial
diffeomorphism}. As in
\cite{Horava:2009uw,Horava:2008ih} we
extend these symmetries to the
diffeomorphisms that respect the
preferred codimension-one foliation
$\mathcal{F}$ of the theory by the
slices of fixed time. By definition
such a foliation-preserving
diffeomorphism consists a space-time
dependent spatial diffeomorphisms as
well as time-dependent time
reparameterization. These symmetries
are now generated by infinitesimal
transformations
\begin{equation}\label{fpd}
\delta x^i\equiv
x'^i-x^i=\zeta^i(t,\bx) \ , \quad
\delta t\equiv t'-t=f(t) \ .
\end{equation}
It was shown in \cite{Horava:2008ih}
that the metric transform under
(\ref{fpd}) as
\begin{eqnarray}\label{trm}
g'_{ij}(t',\bx')&=&g_{ij}(t,\bx)-
g_{il}(t,\bx)\partial_j
\zeta^l(t,\bx)-\partial_i
\zeta^k(t,\bx) g_{kj}(t,\bx) \ .  \nonumber \\
%g'^{ij}(t',\bx')&=& g^{ij}(t,\bx)+
%\partial_n \zeta^i(t,\bx) g^{nj}(t,\bx)
%+g^{in}(t,\bx)
%\partial_n \zeta^j(t,\bx)
% \ . \nonumber \\
\end{eqnarray}
The original action (\ref{SHGG}) is not
invariant under (\ref{fpd}). On the
other hand it was shown in
\cite{Horava:2008ih} that in order to
find an action that is invariant under
(\ref{fpd}) it is necessary to
introduce new fields $N_i(t,\bx),N(t)$
that transform under (\ref{fpd}) as
\begin{eqnarray}
N'_i(t',\bx')&=& N_i(t,\bx) -N_i(t,\bx)
\dot{f}(t)-N_j(t,\bx)\partial_i
\zeta^j(t,\bx)-g_{ij}(t,\bx)
\dot{\zeta}^j(t,\bx) \ ,   \nonumber \\
% N'^i(t',\bx')
%&=&N^i(t,\bx)+N^j(t,\bx)\partial_j
%\zeta^i(t,\bx)-
%N^i(t,\bx)\dot{f}(t)-\dot{\zeta}^i(t,\bx)
%\ ,
\nonumber \\
N'(t')&=&N(t)-N(t) \dot{f}(t) \ .
 \nonumber \\
\end{eqnarray}
As the next step we have to replace
volume element $dt d^D\bx \sqrt{g}$
with  $dt d^D\bx N\sqrt{g}$ and  the
time derivative of $g_{ij}$ with
\begin{equation}
\partial_t g_{ij}\Rightarrow 2K_{ij} \
,
\end{equation}
where $K_{ij}$ is defined as
\begin{equation}\label{Kdef}
K_{ij} =\frac{1}{2N}(\partial_t g_{ij}-
\nabla_i N_j-\nabla_j N_i)  \ ,
\end{equation}
and where $\nabla_i$ is $D$ dimensional
covariant derivative constructed from
the metric components $g_{ij}$. It can
be shown that (\ref{Kdef})  transform
covariantly under (\ref{fpd})
%. In fact
%\begin{eqnarray}
%\frac{1}{N'}\partial_{t'}
%g'_{ij}(t',\bx')=\frac{1}{N(t,\bx)}
%\partial_t g_{ij}(t,\bx)+\frac{1}{N}
%\partial_t g_{ij}\dot{f}
%-\frac{1}{N}\partial_t g_{ij}\dot{f}
%-\frac{1}{N}\partial_k g_{ij}\partial_t \zeta^k
%\nonumber \\
%\end{eqnarray}
%and hence
\begin{eqnarray}
K'_{ij}(t',\bx')=
K_{ij}(t,\bx)-K_{ik}(t,\bx)\partial_j
\zeta^k(t,\bx)
-\partial_i\zeta^k(t,\bx) K_{kj}(t,\bx)
 \ .
\nonumber \\
\end{eqnarray}
Performing these substitutions in
 (\ref{SHGG}) we find the gravity
action invariant under the foliation
preserving diffeomorphism in the form
\begin{eqnarray}\label{SHGGfpd}
S=-\kappa^2 \int dt d^D\bx
\sqrt{g}N\left( \sqrt{1+\beta E^{ij}
\mG_{ijkl} E^{kl}}
\sqrt{1-\frac{4}{\kappa^4\beta}
(K_{ij}K^{ij} -\lambda K^2)}-1\right) \
. \nonumber
\\
\end{eqnarray}
Note also that linearized form of the
action (\ref{SHGGfpd}) takes the form
\begin{eqnarray}
S=\frac{1}{2} \int dt d^D\bx
\sqrt{g}N\left(\frac{4}{\kappa^2\beta}
(K_{ij}K^{ij}-\lambda K^2)-
\kappa^2\beta E^{ij} \mG_{ijkl} E^{kl}
\right) \  \nonumber
\\
\end{eqnarray}
that after trivial rescaling of
parameter $\beta$ resembles the
Ho\v{r}ava's form of the gravity
theory. For that reason we can consider
the action (\ref{SHGGfpd}) as the
$f(R)$-like version of the
Hora\v{r}ava-Lifshitz gravity.

In the next subsection  we develop the
Hamiltonian formalism of given theory.
%%%%%%%%%%%%%%%%%%%%%%%%%%%%%%%%%%%%%%%%%%%555
\subsection{Hamiltonian Formalism}
The dynamical variables of the theory
are $N_i(\bx),\pi^i(\bx),N,\pi^N$
together with
$g_{ij}(\bx),\pi^{ij}(\bx)$ with
corresponding non-zero Poisson brackets
\begin{equation}
\pb{g_{ij}(\bx),\pi^{kl}(\by)}=
\frac{1}{2}(\delta_i^k\delta_j^l+
\delta_i^l\delta_j^k)\delta(\bx-\by) \
, \quad \pb{N^i(\bx),\pi_j(\by)}=
\delta^i_j\delta(\bx-\by) \ , \quad
\pb{N,\pi^N}=1 \ .
\end{equation}
Note that  $N(t)$ and $\pi^N(t)$ are
homogeneous functions of time. In other
words they obey projectability
condition which has an important
consequence for the consistency of the
Ho\v{r}ava-Lifshitz theory
\cite{Mukohyama:2009tp}. Further, as
follows from the form of the action
(\ref{SHGGfpd}) the  momenta $\pi^{ij}$
conjugate to $g_{ij}$ can be expressed
as function of $g_{ij}$ and
$\partial_tg_{ij}$ from the relation
\begin{eqnarray}\label{defpi}
\pi^{ij}(\bx)= \frac{\delta S}{\delta
\partial_t g_{ij}(\bx)}=
\frac{2}{\kappa^2\beta}\frac{\sqrt{g}
\mG^{ijkl}K_{kl}}
{\sqrt{1-\frac{4}{\beta\kappa^4}
K_{ij}\mG^{ijkl}K_{kl}}} \sqrt{ 1+\beta
E^{ij}\mG_{ijkl}E^{kl}}   \ .
 \nonumber \\
%\frac{4}{\beta\kappa^4}K_{ij}
%\mG^{ijkl}K_{kl}
%=\frac{1}{(\sqrt{g})^2}\beta
%\pi^{ij}\mG_{ijkl}\pi^{kl}
%\frac{1}{\frac{1}{(\sqrt{g})^2}\beta
%\pi^{ij}\mG_{ijkl} \pi^{kl}+ (\sqrt{
%\alpha+\frac{\beta}{4\kappa^4_W}
%(R^{ij}-\frac{1}{2}g^{ij}) \mG_{ij,kl}
%(R^{kl}-\frac{1}{2}g^{kl}) })^2} \ .
%\nonumber \\
\end{eqnarray}
On the other hand since  the time
derivative of $N_i$ and $N$ do not
appear in the action (\ref{SHGGfpd})
 we find that
the momenta $\pi^i$ and $\pi^N$ form
the primary constraints of the theory
\begin{equation}
\pi^i(t,\bx)\approx 0 \ , \quad
\pi^N(t)\approx 0 \ .
\end{equation}
Finally the standard definition of the
Hamiltonian density gives
\begin{eqnarray}
\mH&=&\partial_t g_{ij}\pi^{ij}-\mL=
\nonumber \\
%=\kappa^2 \sqrt{g}N \frac{1}{
%\sqrt{1-\frac{4}{\kappa^2}K_{ij}
%\mG^{ijkl}K_{kl}}} \sqrt{
%\alpha+\frac{\beta}{4\kappa^4_W}
%(R^{ij}-\frac{1}{2}g^{ij}) \mG_{ij,kl}
%(R^{kl}-\frac{1}{2}g^{kl})
%}-\kappa^2\sqrt{\alpha}+\nonumber \\
%+\frac{2}{\kappa^2\beta} \sqrt{g}\frac{
%(\nabla_i N_j+ \nabla_j
%N_i)\mG^{ijkl}K_{kl}} {
%\sqrt{1-\frac{4}{\kappa^2}K_{ij}
%\mG^{ijkl}K_{kl}}} \sqrt{
%\alpha+\frac{\beta}{4\kappa^4_W}
%(R^{ij}-\frac{1}{2}g^{ij}) \mG_{ij,kl}
%(R^{kl}-\frac{1}{2}g^{kl}) }= \nonumber
%\\
&=&\kappa^2 \sqrt{g}N \left(\sqrt{1+
\frac{1}{g}\pi^{ij}\mG_{ijkl}\pi^{kl}+
\beta E^{ij}\mG_{ijkl}E^{kl}}-1\right)+
\nonumber \\
% (R^{ij}-\frac{1}{2}g^{ij})
%\mG_{ij,kl}
%(R^{kl}-\frac{1}{2}g^{kl})}+ \nonumber
%\\
&+& (\nabla_i N_j+\nabla_j N_i)\pi^{ij}
\ .
 \nonumber \\
\end{eqnarray}
As a consequence we find that the
Hamiltonian is equal to
\begin{eqnarray}
H&=&\int d^D\bx \mH= \int d^D \bx (N(t)
\mH_0(\bx,t)+ N_i(t,\bx)\mH^i(\bx,t)) \ , \nonumber \\
\mH_0&=&\kappa^2 \sqrt{g}\left(\sqrt{1+
\frac{1}{g}\pi^{ij}\mG_{ijkl}\pi^{kl}+
E^{ij}\mG_{ijkl}E^{kl}}
%\frac{\beta}{4\kappa^4_W}
%(R^{ij}-\frac{1}{2}g^{ij}) \mG_{ij,kl}
%(R^{kl}-\frac{1}{2}g^{kl})}
-1\right)
%\equiv  \nonumber \\
%\equiv \kappa^2 \sqrt{g}\mH_0' \ ,
\ , \quad \mH^i=-2\nabla_j \pi^{ij} \ ,
\nonumber
\\
\end{eqnarray}
where we ignore boundary terms.

 The
primary constraints $\pi^i(\bx)\approx
0 \ , \pi^N(t)\approx 0$ have to be
preserved during the time evolution of
the system and consequently
\begin{equation}
\partial_t
\pi^i(\bx)=\pb{\pi^i(\bx),H}=-\mH^i(\bx)\approx
0
 \ , \quad
 \partial_t \pi^N(t)=\pb{N(t),H}=-\int d^D\bx\mH_0(\bx)\approx
 0 \ .
 \end{equation}
Since the right side of the equations
above have to vanish on constraint
surface we find that the consistency of
the primary constraints generate the
secondary ones
\begin{equation}
\mH^i(\bx)\approx 0 \ , \quad \bT_T\equiv \int d^D\bx
\mH_0(\bx)\approx 0 \ .
\end{equation}
It is convenient to introduce the
smeared form of the diffeomorphism constant
$\bT_S$ defined as
\begin{eqnarray}\label{bT}
\bT_S(\zeta)&=&
\int d^D\bx \zeta_i(\bx) \mH^i(\bx) \ .
\nonumber \\
% \bT_T(f)&=&\int d^D \bx f(t)\mH_0(\bx)=
% f(t)\bT_T \ ,  \quad \bT_T=\int d^D\bx \mH_0(\bx) \ .
%  \nonumber \\
\end{eqnarray}
The next goal is to calculate the Poisson
bracket of constraints $\bT_T$ and
$\bT_S$. Trivially we have that
\begin{eqnarray}
\pb{\bT_T,\bT_T}=0 \ .
%f(t)g(t)\pb{\bT_T,\bT_T}=0 \ . \nonumber \\
\end{eqnarray}
Now we calculate the Poisson brackets
between $\bT_S(\zeta)$ and $\bT$
\begin{eqnarray}
\pb{\bT_S(\zeta),\bT_T}
% \int d^D
%\bx f(t) \pb{\bT(\zeta),\pi^{ij}(\bx)}
%\frac{\delta \mH_0}{\delta
%\pi^{ij}(\bx)}+ \nonumber \\
%+\int d^D\bx f(t)\pb{\bT(\zeta),
%g^{ij}(\bx)}\frac{\delta \mH_0}{\delta
%g^{ij}(\bx)}=
% \nonumber \\
%=\int d^D\bx( f(t)
%[-\kappa^2\zeta^k\partial_k (\sqrt{\det
%g})-\partial_i\zeta^i\sqrt{\det g}]
%\mH_0'-\kappa^2\sqrt{g}
%\zeta^k\partial_k\mH'_0)=
%\nonumber \\
%=-\int d^D\bx
%f(t)(\zeta^k\partial_k\mH_0+\partial_k\zeta^k
%\mH_0)=\nonumber \\
=-\int d^D\bx(
\zeta^k\partial_k\mH_0-\partial_k(
\mH_0)\zeta^k )=0 \ ,
% \int d^D\bx \partial_k
% f(t)\mH_0\zeta^k=0 \ ,
\nonumber \\
\end{eqnarray}
where we used the Poisson bracket
between $\bT_S(\zeta)$ and $\mH_0$
\footnote{For more detailed
calculation, see (\ref{bTSmh0}).}
\begin{eqnarray}
\pb{\bT_S(\zeta),\mH_0}
 &=&-\partial_k\zeta^k \mH_0-
 \zeta^k\partial_k \mH_0 \ .
\nonumber \\
\end{eqnarray}
% using
% \begin{eqnarray}
% \pb{\bT(\zeta),g_{ij}(\bx)}&=& -\zeta^k
% \partial_k
% g_{ij}-g_{jk}\partial_i\zeta^k- g_{ik}
% \partial_j \zeta^k \ , \nonumber \\
% \pb{\bT(\zeta),\pi^{ij}(\bx)}&=&
% -\partial_k (\pi^{ij}\zeta^k) +\pi^{jk}
% \partial_k\zeta^i+\pi^{ik}\partial_k
% \zeta^j \ . \nonumber \\
% \end{eqnarray}
Finally we calculate the Poisson
bracket
\begin{eqnarray}
\pb{\bT_S(\zeta),\bT_S(\xi)}=
\bT_S(\zeta^i\partial_i\xi-\xi^i\partial_i\zeta)
\ .
 \nonumber \\
\end{eqnarray}
In summary we find that the
algebra of constraints for generalized
Ho\v{r}ava-Lifshitz theory that respects
the projectability condition takes very simple
form
\begin{eqnarray}
\pb{\bT_T,\bT_T}&=&0 \ , \nonumber \\
\pb{\bT_S(\zeta),\bT_T}&=&0 \ ,
\nonumber
\\
\pb{\bT_S(\zeta),\bT_S(\xi)}&=&
\bT_S(\zeta^i\partial_i\xi-\xi^i\partial_i\zeta)
\ .
\nonumber \\
\end{eqnarray}
The fact that the algebra of
constraints is closed
 for any theory of gravity that
  obeys the projectability
condition is very attractive. This
result in contrast with the situation
of gravity without the projectability
condition when the algebra is not
closed and the structure constants of
the theory depend on phase space
variables. On the other hand there are
still many unsolved problems and issues
considering Ho\v{r}ava-Lifshitz gravity
theories as was reviewed carefully in
\cite{Mukohyama:2009tp,Blas:2009yd} so
that these results should be taken with
great care.
% In other words the constraint algebra
% of the gravity action that is invariant
% under foliation preserving
% diffeomorphism is very simple.
%  On the
% other hand it is an open question
% whether this algebra allows any
% physical excitations at all.
%%%%%%%%%%%%%%%%%%%%%%%%%%%%%%%%%%%%%%%%%%%%%%%%%%%
\section{ Ho\v{r}ava-Lifshitz $f(R)$  Theory of
Gravity-Without  Projectability
Condition}\label{fourth} In this
section we address the question of the formulation
of the local  form of the
 \emph{condition of detailed
balance}. We again start  with the
assumption that one can define
 the vacuum wave functional
of $D+1$-dimensional quantum theory and
that this functional has the  form as
in (\ref{vvf}). Now we demand that this
vacuum wave functional is annihilated by
%\begin{equation}
%\Psi_0[g(\bx)]=\exp[-\frac{1}{2}W] \ ,
%\end{equation}
%where the explicit form of $W$ was
%given
%
%
% is a functional of
%$D-$dimensional metric and where we
%presume that $W$ is invariant under
%\emph{spatial diffeomoprhism}
%%$x'^i=x'^i(\bx)$.
% The formal norm of
%this functional is again equal to the
%partition function of $D$-dimensional
%gravity theory.
\begin{equation}\label{defH}
\hat{H}=\int d^D\bx \left(N(t,\bx)
\hat{\mH}_0(t,\bx)+N_i(t,\bx)
\hat{\mH}^i(t,\bx)\right) \ ,
\end{equation}
where $\hat{\mH^i}$ is the generator of
spatial diffeomorphism
\begin{equation}
\hat{\mH}^i(\bx)=-2\hat{\nabla}_j\hat{\pi}^{ji}(\bx)
\ ,
\end{equation}
and where  we assume that
$\hat{\mH}_0$
can be written as
\begin{equation}
\hat{\mH}_0(\bx)= \kappa^2\sqrt{\hat{g}}\left(
\sum_{n=0}^\infty
\hc_n(\hat{g}_{ij}) (\hat{Q}^{\dag ij}
\frac{1}{\hat{g}}\hat{\mG}_{ijkl}
\hat{Q}^{kl})^n-\hat{c}_0(\hat{g}_{ij})\right) \ ,
\end{equation}
where $\hat{Q}^{ij}, \hat{Q}^{\dag ij}$
were defined in (\ref{hQij}) and the
functional form of $\hat{E}^{ij}$
follows from (\ref{defE}).
% where again we
%introduced "creation" and
%"annihilation" operators
%\begin{equation}
%\hat{Q}^{\dag ij}=-i\hat{\pi}^{ij}+
%\sqrt{\hat{g}}\hat{E}^{ij} \ , \quad
%\hat{Q}^{\dag ij}=i\hat{\pi}^{ij}+
%\sqrt{\hat{g}}\hat{E}^{ij} \ . \quad
%\end{equation}
%Observe that these operators take
%following form in Schr\"{o}dinger
%representation
%\begin{equation}
%\hat{Q}^{ij}(\bx)=-\frac{\delta}
%{\delta
%g^{ij}(\bx)}+\sqrt{g}(\bx)E^{ij}(\bx) \
%, \quad \hat{Q}^{\dag ij}(\bx)=
%\frac{\delta }{\delta  g^{ij}(\bx)}+
%\sqrt{g}(\bx)E^{ij}(\bx) \ .
%\end{equation}
%Note also that $E^{ij}$ are defined from
%the $D$-dimensional functional $W$ as
%\begin{equation}
%\sqrt{g}E^{ij}(\bx)=\frac{1}{2}\frac{
%\delta W}{\delta g^{ij}(\bx)} \ .
%\end{equation}
%Then it is clear that
%\begin{equation}
%\hat{Q}^{ij}(\bx)\Psi_0[g(\bx)]=
%-\frac{1}{2}\frac{\delta W}{\delta g^{ij}(\bx)}
%+\frac{1}{2}\frac{\delta W}{\delta g^{ij}(\bx)}=0
%\end{equation}
Then it is obvious that the local
constraint $\hat{\mH}_0(\bx)$
annihilates the vacuum wave functional
(\ref{vvf}). Since  $W$
is invariant under spatial
diffeomorphism by construction
 we find that the vacuum
wave functional $\Psi_0$ is annihilated
by $\hat{H}$  (\ref{defH}) as well.
 We should
again stress the important fact that
(\ref{defH}) contains the lapse
function that depends on $\bx$ as well.
Note that we only  demand that this
Hamiltonian annihilates the vacuum
state functional while its action on
other states of the theory is not
specified. This is different from the
standard constraint of general
relativity where the Dirac analysis
implies that \emph{all} wave
functionals should be annihilated by
Hamiltonian and diffeomorphism
constraints. On the other hand we will
see below that the correct Hamiltonian
treatment of the theory specified by
the Hamiltonian above will lead to the
requirement that all states should be
annihilated by (\ref{defH}).

As usual we are interested in the
Lagrangian formulation of given theory.
In order to find it we  consider the
classical form of the Hamiltonian
(\ref{defH}) and we also restrict
ourselves to the  following
 form of the Hamiltonian density $\mH_0$
\begin{eqnarray}\label{mH0}
\mH_0=\kappa^2 \sqrt{g}\left(\sqrt{1+
\beta
(-i\pi^{ij}+\sqrt{g}E^{ij})\frac{1}{g}
\mG_{ijkl}
(i
\pi^{kl}+ \sqrt{g}\frac{1}{2}E^{kl})}-1\right) \ . \nonumber \\
\end{eqnarray}
% Note that   $E^{ij}$ transform
% under spatial diffeomorphisms as
% \begin{equation}
% E'^{ij}(\bx')=E^{kl}(\bx)D_k^i D_l^j \ ,
% \end{equation}
% or when  $x'^i=x^i+\zeta^i \ , D^i_j=
% \delta^i_j+\partial_i\zeta^j \ ,
% \iD^i_j= \delta^i_j-\partial_i\zeta^j$
% so that
% \begin{eqnarray}
% E'^{ij}(\bx')=E'^{ij}(\bx)+
% \partial_k E^{ij}(\bx)\zeta^k=
% E^{ij}(\bx)+ E^{il}\partial_l\zeta^j+
% \partial_k\zeta^iE^{kj} \ .
% \nonumber \\
% \end{eqnarray}
% Further, if we introduce the metric
% $\mG_{ijkl}$ in the form
% \begin{equation}
% \mG_{ijkl}=\frac{1}{2}(g_{ik}g_{jl}+
% g_{il}g_{jk})-\tilde{\lambda}g_{ij}g_{kl}
% \end{equation}
% that under infinitesimal form of
% spatial diffeomorphisms transform as
% \begin{eqnarray}
% \mG'_{ijkl}(\bx')= \frac{1}{2} \left(
% (g_{ik}+\triangle
% g_{ik})(g_{jl}+\triangle g_{jl})
% +(g_{il}+\triangle g_{il})( g_{jk}
% +\triangle g_{jk})\right)- \nonumber \\
% -\tilde{\lambda} (g_{ij}+\triangle
% g_{ij})(g_{kl}+\triangle g_{kl})
% =\mG_{ijkl}(\bx)-
% \partial_i \zeta^m \mG_{mkjl}-
% \mG_{imjl}\partial_k \zeta^m-
% \mG_{ijml}
% \partial_k \zeta^m-\mG_{ijkm}
% \partial_l \zeta^l
% \nonumber \\
% \delta \mG_{ijkl}(\bx)= \mG'_{ijkl}
% (\bx)-\mG_{ijkl}(\bx)= -\partial_m
% \mG_{ijkl}(\bx)-
% \partial_i \zeta^m \mG_{mkjl}-
% \mG_{imjl}\partial_k \zeta^m-
% \mG_{ijml}
% \partial_k \zeta^m-\mG_{ijkm}
% \partial_l \zeta^l \nonumber \\
% \end{eqnarray}
% that implies that $E^{ij} \mG_{ijkl}
% E^{kl}$ is invariant under spatial
% diffeomorphisms.
  Then using
 (\ref{defH}) and  (\ref{mH0})
we find that the  time derivative of
$g_{ij}$ is equal to
 \begin{eqnarray}\label{partg}
 \partial_t g_{ij}=
 \pb{g_{ij},H}=
 \kappa^2 \frac{N\beta}{\sqrt{g}}
 \frac{\mG_{ijkl}\pi^{kl}}
 {\sqrt{1+\frac{\beta}{g}\pi^{ij}\mG_{ijkl}
 \pi^{kl}+\beta E^{ij}\mG_{ijkl}E^{kl}}}+
 \nabla_j N_i+\nabla_i N_j \ ,
\nonumber \\
\end{eqnarray}
where we used the canonical Poisson
brackets
\begin{equation}
\pb{g_{ij}(\bx),\pi^{kl}(\by)}=
\frac{1}{2}(\delta_i^k\delta_j^l+
\delta_i^l\delta_j^k) \delta(\bx-\by) \
\end{equation}
and the fact that
\begin{eqnarray}
& &\pb{g_{ij}(\bx),\int d^D\by
N_k(\by)\mH^k(\by)}=
 -2\int d^D\by N_k(\by)
\nabla_{l}(\pb{g_{ij}(\bx),\pi^{kl}(\by)})=
\nonumber \\
% -\int d^D\bx N_k(\by)
%\nabla_l
%((\delta_i^k\delta_j^l+\delta_i^l\delta_j^k)
%\delta(\bx-\by))=
%\nonumber \\
&=&\nabla_j N_i(\bx)+\nabla_i N_j(\bx)
\ .
\nonumber \\
\end{eqnarray}
The equation (\ref{partg}) implies that
it is natural to introduce the tensor
$K_{ij}= \frac{1}{2N}(\partial_t
g_{ij}- \nabla_j N_i-\nabla_i N_j)$ so
that (\ref{partg}) can be written as
\begin{eqnarray}
 %\frac{1}{N}(\partial_t g_{ij}-\nabla_i
% N_j-\nabla_j N_i)= \kappa^2 \frac{\beta}{\sqrt{g}}
% \frac{\mG_{ijkl}\pi^{kl}}
% {\sqrt{1+\frac{\beta}{g}\pi^{ij}\mG_{ijkl}
% \pi^{kl}+\beta E^{ij}\mG_{ijkl}E^{kl}}}
% \Rightarrow \nonumber \\
 2K_{ij}= \kappa^2 \frac{\beta}{\sqrt{g}}
 \frac{\mG_{ijkl}\pi^{kl}}
 {\sqrt{1+\frac{\beta}{g}\pi^{ij}\mG_{ijkl}
 \pi^{kl}+ \beta
 E^{ij}\mG_{ijkl}E^{kl}}} \ .
 \nonumber \\
\end{eqnarray}
Clearly using this relation
 we can express $\pi^{ij}$ as
a function of $g_{ij},K_{ij}$. Then
after some algebra we find
%
%This equation can be easily inverted so
%that
%\begin{eqnarray}
% \frac{\beta}{g}\pi^{ij}\mG_{ijkl}
% \pi^{kl}=4\frac{K_{ij}\mG^{ijkl}K_{kl}(1+
% \beta E^{ij}\mG_{ijkl}
% E^{kl})}{
% (\kappa^4\beta-4K_{ij}\mG^{ijkl}
% K_{kl})} \ . \nonumber \\
% \end{eqnarray}
the Lagrangian in the form
\begin{eqnarray}\label{LgH}
L&=&\int d^D\bx(\partial_t
g_{ij}\pi^{ij}-N\mH_0-N_i\mH^i)=
\nonumber \\
%=\int d^D\bx (- N\kappa^2\sqrt{g}
%\frac{\alpha+\beta
%E_{ij}\mG^{ijkl}E_{kl}}{ \sqrt{\alpha+
%\frac{\beta}{g}\pi^{ij}\mG_{ijkl}\pi^{kl}+
%\beta
%E^{ij}\mG_{ijkl}E^{kl}}}+N\sqrt{g}\kappa^2\sqrt{\alpha})=
%\nonumber \\
&=&-\kappa^2\int d^D\bx \sqrt{g}N
\left(\sqrt{1+\beta E^{ij}\mG_{ijkl}
E^{kl}}\sqrt{1-\frac{4}{\kappa^4\beta}
K_{ij}\mG^{ijkl}K_{kl}}-1\right) \ .
\nonumber \\
\end{eqnarray}
We see that this Lagrangian
takes completely the same form as
the Lagrangian given in (\ref{SHGGfpd}).
 However it is crucial
that in the new formulation the field
$N$ depends on $\bx$ and $t$ as well.
In other words we derived the
Ho\v{r}ava-Lifshitz $f(R)$ gravity theory
without projectability condition.

\subsection{Hamiltonian Formalism}
We see that  the Lagrangian density
(\ref{LgH}) depends on $N(t,\bx)$ and
$N^i(t,\bx)$ that can be interpreted
as additional fields in
the theory. Then when we proceed
to the Hamiltonian formalism we find that
the phase space of the theory is spanned
by $N,N^i$ with conjugate momenta $\pi_N,\pi_i$
and metric components $g_{ij}$ with conjugate
momenta $\pi^{ij}$.  The fact that
the Lagrangian (\ref{LgH}) does not contain
time derivatives of $N$ and $N^i$
implies that the  momenta
$\pi^i(\bx),\pi_N(\bx) $ vanish and
form the primary constraints of the
theory. Finally the standard analysis
of constraints system implies
that the Hamiltonian (\ref{defH}) with
$\mH_0$ given in (\ref{mH0}) is a
sum of the local constraints
\begin{equation}
\mH_0(\bx)\approx 0 \ ,
\mH^i(\bx)\approx 0 \ .
\end{equation}
The quantum mechanical analogue of
these constraints is the requirement
that  \emph{all wave functionals}
should be annihilated by them. Observe
that this is more stronger requirement
then the formulation of the local
balance condition given in the first
paragraph of this section.
 In summary, the consistency
of the theory defined by (\ref{LgH})
implies that at the classical level
the Hamiltonian (\ref{defH})
should be sum of local
constraints. The quantum mechanical
formulation is that all wave
functionals should be annihilated by
the quantum  Hamiltonian (\ref{defH})
again with $\hat{\mH}_0$ given
in (\ref{mH0}).

Now we start to study the algebra
of constraints $\mH^i,\mH_0$ when
we presume the most general form
of the constraint  $\mH_0$
\begin{eqnarray}
\mH_0&=&\kappa^2\sqrt{g}\left(\sum_{n=0}^\infty
c_n(g_{ij}) \left(Q^{\dag
ij}\frac{1}{g}\mG_{ijkl}
Q^{kl}\right)^n-c_0(g_{ij})\right)=\nonumber
\\
&=& \kappa^2 \sqrt{g} \sum_{n=1}^\infty
c_n(g_{ij}) \left (Q^{\dag
ij}\frac{1}{g}\mG_{ijkl} Q^{kl}\right)^n \
.  \nonumber \\
\end{eqnarray}
If we introduce the smeared form of
the diffeomorphism constraint
$\bT_S(\zeta)=\int d^D\bx \zeta_i(\bx)
\mH^i(\bx)$ we can easily determine
Poisson
brackets
\begin{eqnarray}\label{pbhelp}
\pb{\bT_S(\zeta),g_{ij}}&=& -\zeta^k
\partial_k
g_{ij}-g_{jk}\partial_i\zeta^k- g_{ik}
\partial_j \zeta^k \ , \nonumber \\
\pb{\bT(\zeta),\pi^{ij}}&=&
-\partial_k (\pi^{ij}\zeta^k) +\pi^{jk}
\partial_k\zeta^i+\pi^{ik}\partial_k
\zeta^j \ , \nonumber \\
\pb{\bT_S(\zeta),\sqrt{g}}&=&
-\zeta^k\partial_k
\sqrt{g}-\partial_k\zeta^k
\sqrt{g} \ , \nonumber \\
\pb{\bT_S(\zeta),\frac{1}{2}\frac{\delta
W}{\delta g^{ij}}}&=&
%\pb{\bT(\zeta),
%\sqrt{g}E^{ij}(\bx)}= \nonumber \\
% =-\zeta^k\partial_k\sqrt{g}E^{ij}(\bx)
% -\partial_k\zeta^k\sqrt{g}E^{ij}(\bx)-
% \zeta^k\partial_k E^{ij}(\bx)\sqrt{g}+
% \nonumber \\
% +\sqrt{g}E^{ik}\partial_k\zeta^j(\bx)+
% \partial_k \zeta^i E^{kj}(\bx)=
% \nonumber \\
-\partial_k\left(\frac{1}{2}\zeta^k
\frac{\delta W}{\delta
g_{ij}}\right)+ \frac{1}{2}\frac{
\delta W}{\delta g_{ik}}\partial_j
\zeta^k+
\partial_i \zeta^k\frac{1}{2} \frac{\delta W}{\delta
g_{kj}}  \ . \nonumber \\
\end{eqnarray}
 Then it is easy to find that
\begin{eqnarray}\label{bSQ}
\pb{\bT_S(\zeta),Q^{ij}}&=&
% i\pb{\bT(\zeta),\pi^{ij}(\bx)}+
% \frac{1}{2} \pb{\bT(\zeta),\frac{\delta
% W}{\delta g_{ij}(\bx)}}=\nonumber \\
% &=&
-\partial_k (Q^{ij}\zeta^k )+
\partial_k\zeta^i Q^{kj}+
Q^{ik}\partial_k \zeta^j \ ,
\nonumber \\
 \pb{\bT_S(\zeta),Q^{\dag
ij}}&=&
-\partial_k\left(\zeta^k
 Q^{\dag ij}\right)+
\partial_k\zeta^i Q^{ \dag kj}+Q^{\dag
ik}\partial_k\zeta^j \ .
\nonumber \\
\end{eqnarray}
For further purposes we also determine
following Poisson bracket
\begin{eqnarray}
\pb{\bT_S(\zeta), \frac{1}{g}
\mG_{ijkl}}= (2\partial_k\zeta^k
(g)+\zeta^k\partial_k (g))
\frac{1}{g^2}\mG_{ijkl}-
\nonumber \\
-\frac{1}{g}(\partial_p
\mG_{ijkl}\zeta^p+
\partial_i \zeta^p\mG_{pjkl}+
\partial_j\zeta^p\mG_{ipkl}+
\mG_{ijpl}\partial_k\zeta^p+ \mG_{ijkp}
\partial_j\zeta^p) \ .
\nonumber \\
\end{eqnarray}
Then it is easy to see
\begin{eqnarray}
\pb{\bT_S(\zeta), Q^{\dag
ij}\frac{1}{g}\mG_{ijkl}
Q^{kl}}=-\partial_m \left(Q^{\dag
ij}\frac{1}{g}
\mG_{ijkl}Q^{kl}\right)\zeta^m \ .
\nonumber \\
\end{eqnarray}
Using this result and also  the
third equation in (\ref{pbhelp}) we find
\begin{eqnarray}\label{bTSmh0}
\pb{\bT_S(\zeta),\mH_0}
%  \kappa^2
% \pb{\bT(\zeta),\sqrt{g}(\bx)}
% \sum_{n=1}^\infty c_n (Q^{\dag
% ij}\frac{1}{g}\mG_{ijkl}
% Q^{kl})^n+\nonumber \\
% \kappa^2\sqrt{g}\sum_{n=1}^\infty n
% \pb{\bT(\zeta), (Q^{\dag
% ij}\frac{1}{g}\mG_{ijkl}
% Q^{kl})}c_n(Q^{\dag
% ij}\frac{1}{g}
% \mG_{ijkl}
% Q^{kl})^{n-1}=\nonumber \\
&=&\kappa^2[-\zeta^k\partial_k
\sqrt{g}-\partial_k\zeta^k \sqrt{g}]
\sum_{n=1}^\infty c_n \left(Q^{\dag
ij}\frac{1}{g}\mG_{ijkl}
Q^{kl}\right)^n-\nonumber \\
&-&\kappa^2\sqrt{g}\sum_{n=1}^\infty c_n
\partial_m \left(Q^{\dag ij}
\frac{1}{g}\mG_{ijkl}Q^{kl}\right)\zeta^m
\left(Q^{\dag ij}\frac{1}{g}\mG_{ijkl}
Q^{kl}\right)^{n-1}=
 \nonumber \\
 &=&-\partial_k\zeta^k \mH_0-
 \zeta^k\partial_k \mH_0
\nonumber \\
\end{eqnarray}
and when we introduce the smeared form
of the constraint $\mH_0$
\begin{equation}
\bT_T(f)=\int d^D\bx f(\bx)\mH_0(t,\bx)
\end{equation}
we obtain
\begin{eqnarray}
\pb{\bT_S(\zeta),\bT_T(f)}&=&
% \int d^D\by
% f(t,\by)
% \pb{\bT(\zeta),\mH(\by)}=
% \nonumber \\
-\int d^D \bx f(\bx) (\partial_k\zeta^k
\mH_0(\bx)+
 \zeta^k\partial_k \mH_0(\bx))=
\nonumber \\
&=& \int d^D\bx
 \partial_k f(\bx)\zeta^k
 \mH_0=\bT_T(\partial_k f \zeta^k) \ .
\nonumber \\
\end{eqnarray}
Finally  the Poisson brackets of the
diffeomorphism constraints is equal to
\begin{equation}\label{pbsd}
\pb{\bT_S(\zeta),\bT_S(\xi)}=
\bT_S(\zeta^i\partial_i\xi-\xi^i\partial_i\zeta)
\ .
\end{equation}
Now we come to the analysis of the most
intricate Poisson bracket
$\pb{\bT_T(f),\bT_T(\zeta)}$. Note that
the previous Poisson brackets were
valid for any form of the constraint
$\mH_0$. On the other hand we can
certainly find an equivalent constraint
using following observation. The
Hamiltonian constraint has the form
$\mH_0=f(Q^{\dag ij}\mG_{ijkl}Q^{kl})$.
Then instead imposing the constraint
$\mH_0\approx 0$ we can impose the
constraint $\sqrt{g}Q^{\dag
ij}\mG_{ijkl} Q^{kl}\approx 0$. This
fact simplifies the analysis
considerably however it is still very
intricate as was shown for example in
\cite{Li:2009bg} where the analysis of the constraint
algebra  of $3+1$ dimensional
Ho\v{r}ava-Lifshitz theory was
performed with  the result that the
Poisson bracket of the
 constraint $\sqrt{g}Q^{\dag
ij}\mG_{ijkl} Q^{kl}\approx 0$ is not
closed but it generates new additional
ones. The upshot of this analysis is
that  it seems that the resulting
theory does not contain any physical
degrees of freedom. This seems to be a
serious problem  of the
Ho\v{r}ava-Lifshitz theory without
projectability condition. On the other
hand there exists an  alternative
procedure how to solve the constraint
$\mH_0\approx 0$. This idea  was
suggested in the original Ho\v{r}ava
work \cite{Horava:2008ih}. Explicitly,
the form of the constraint
$\mH_0(\bx)\approx 0$
 suggests that the constraint
$\mH_0(\bx)\approx 0$ can be solved by
collection of constraints
$Q^{ij}(\bx)\approx 0$. In other words
we propose following
 alternative  set of constraints of
$f(R)$ Ho\v{ra}va-Lifshitz gravity
\begin{equation}
\mH^i(\bx)\approx 0 \ , \quad
Q^{ij}(\bx)\approx 0 \
\end{equation}
or their smeared form
\begin{equation}\label{Qsm}
\bT_S(\zeta)= \int d^D\bx \zeta_i(\bx)
\mH^i(\bx) \ , \quad
\mathbf{Q}(\Lambda)= \int d^D\bx
\Lambda_{ij}(\bx)Q^{ij}(\bx) \ .
\end{equation}
 Let us now show
that this set of constraints
 forms the closed
algebra. Since
\begin{eqnarray}
\pb{Q^{ij}(\bx),Q^{kl}(\by)}=
%\frac{i}{2}
%\pb{\pi^{ij}(\bx),\frac{\delta
%W}{\delta g^{kl}(\by)}}
%+\frac{i}{2}\pb{\frac{\delta W}{\delta
%g^{ij}(\bx)},\pi^{kl}(\by)}= \nonumber
%\\
-\frac{i}{2} \frac{\delta}{ \delta
g^{ij}(\bx)}\frac{\delta W}{\delta
g^{kl}(\by)}+ \frac{i}{2} \frac{\delta
}{\delta g^{kl}(\by)} \frac{\delta
W}{\delta g^{ij}(\bx)}=0 \nonumber \\
\end{eqnarray}
we easily find that
\begin{equation}
\pb{\mathbf{Q}(\Lambda),
\mathbf{Q}(\Gamma)}=0 \ .
\end{equation}
Further, using (\ref{bSQ}) we find
\begin{eqnarray}
\pb{\bT_S(\zeta),\mathbf{Q}(\Lambda)}&=&
\int d^D \bx (
\partial_k\Lambda_{ij}
Q^{ij}\zeta^k+
\partial_k \zeta^i Q^{kj}+
Q^{ik}\partial_k \zeta^j)= \nonumber \\
&=&\mathbf{Q}(\partial_k
\Lambda_{ij}\zeta^k+
\partial_i\zeta^k\Lambda_{kj}+\Lambda_{ik}
\partial_j \zeta^k) \ . \nonumber \\
\end{eqnarray}
These Poisson brackets together with
(\ref{pbsd}) imply that the algebra of
the constraints
 (\ref{Qsm}) is closed. On the other
 hand as was stressed originally in
 \cite{Horava:2008ih} this set of
 constraints is certainly too strong
 and it turns out that the resulting
 theory is topological without any
 local excitations. This conclusion
 however suggests that the
 Ho\v{r}ava-Lifshitz theory of gravity
 without projectability condition
has natural physical interpretation as
the topological theory of gravity.

%%%%%%%%%%%%%%%%%%%%%%%%%%%%%%%%%%%%%%%%%%%
\subsection{Ultralocal Gravity}
In this section we present an example
of the Ho\v{r}ava-Lifshitz $f(R)$
gravity that has closed algebra of
constraints. Using terminology
introduced in  \cite{Horava:2009uw} we
call this theory as \emph{ultralocal
Ho\v{r}ava-Lifshitz $f(R)$ gravity}.

The simplest example of the ultralocal
theory  is characterized by condition
that
\begin{equation}
E^{ij}=0 \ .
\end{equation}
Since in this case $Q^{ij}=-Q^{\dag ij}$ we
find
% \begin{equation}
% \pb{Q^{ij}(\bx),Q^{kl}(\by)}=
% =\pb{Q^{\dag ij}(\bx),Q^{kl}(\by)}=
% \pb{Q^{\dag ij}(\bx),Q^{kl}(\by)}=0
% \end{equation}
% To proceed we have to determine
% following Poisson brackets
\begin{eqnarray}\label{QQul}
\pb{Q^{\dag ij}(\bx),Q^{kl}(\by)}&=&0 \ , \nonumber \\
\pb{Q^{ij}(\bx),\frac{1}{g}
\mG_{klmn}(\by)}&=&
%  \pb{i\pi^{ij}(\bx),
% \frac{1}{\sqrt{g}}(\by)}\mG_{klmn}(\by)+
% \frac{1}{\sqrt{g}}\pb{i\pi^{ij}(\bx),
% \mG_{ijkl}(\by)}= \nonumber \\
-\pb{Q^{\dag ij}(\bx),
\frac{1}{g}\mG_{klmn}(\by)} \nonumber
\\
\end{eqnarray}
% using the fact that $Q^{\dag
% ij}=-Q^{ij}$. Using this fact we easily
% obtain
and consequently
\begin{eqnarray}
\pb{Q^{\dag ij}\frac{1}{g}
\mG_{ijkl}Q^{kl}(\bx), Q^{\dag mn}
\frac{1}{g}\mG_{mnpq}Q^{pq}(\by)}=0 \ .
\nonumber \\
% =-2\pb{Q^{ij}(\bx)\frac{1}{g}
% \mG_{mnpq}(\by)}Q^{\dag mn}Q^{pq}(\by)
% \frac{1}{q}\mG_{ijkl}Q^{kl}(\bx)
% -\nonumber \\
% -2\pb{\frac{1}{g}\mG_{ijkl}(\by),
% Q^{mn}(\bx)}
% Q^{pq}(\by)\frac{1}{g}\mG_{mnpq}
% Q^{\dag ij}Q^{kl}(\bx)= \nonumber \\
% =-2\pb{Q^{ij}(\bx)\frac{1}{g}
% \mG_{mnpq}(\by)}Q^{\dag mn}Q^{pq}(\by)
% \frac{1}{q}\mG_{ijkl}Q^{kl}(\bx)
% +\nonumber \\
% +2\pb{Q^{ij}(\by)\frac{1}{g}
% \mG_{mnpq}(\bx)}Q^{\dag mn}Q^{pq}(\bx)
% \frac{1}{q}\mG_{ijkl}Q^{kl}(\by)=0
% \nonumber \\
\end{eqnarray}
% using the fact that $\pb{Q^{ij}(\bx),
% \frac{1}{g}\mG_{klmn}(\by)}\sim
% \delta(\bx-\by)$. This result has an
% important consequence since now
Then it is easy to determine the Poisson
brackets of the constraints $\mH_0\approx 0$
using the fact that
\begin{eqnarray}\label{mH0ul}
\pb{\mH_0(\bx),\mH_0(\by)}&=& \int
d\bx' d\by' \frac{\delta
\mH_0(\bx)}{\delta (Q^{\dag
ij}\mG_{ijkl} Q^{kl})(\bx')} \pb{
(Q^{\dag ij}\mG_{ijkl}Q^{kl})(\bx'),
(Q^{\dag
ij}\mG_{ijkl}Q^{kl})(\by')}\times
\nonumber \\
&\times & \frac{\delta
\mH_0(\by)}{\delta (Q^{\dag
ij}\mG_{ijkl} Q^{kl})(\by')}=0 \ .
\nonumber \\
% \beta^2 \frac{1}{\sqrt{\alpha+ \beta
% (Q^{\dag ij}\mG_{ijkl} Q^{kl})}(\bx)}
% \pb{(Q^{\dag ij}\mG_{ijkl}Q^{kl})(\bx),
% (Q^{\dag
% ij}\mG_{ijkl}Q^{kl}(\by)}\times
% \nonumber \\
% \times \frac{1}{\sqrt{\alpha+ \beta
% (Q^{\dag ij}\mG_{ijkl} Q^{kl})}(\by)}=0
% \nonumber \\
\end{eqnarray}
Let us now consider the second
 example of ultralocal
theory  when $W$ has the form
\begin{equation}
W=\Lambda \sqrt{g} \ .
\end{equation}
In this case we easily find
\begin{equation}
Q^{\dag ij}=-i\pi^{ij}+
\frac{1}{4}\Lambda g^{ij}\sqrt{g} \ ,
\quad Q^{ij}= i\pi^{ij}+\frac{1}{4}
\Lambda g^{ij}\sqrt{g} \ .
\end{equation}
Now the Poisson brackets between
$Q^{\dag ij}$ and $Q^{kl}$ is non-zero
%\begin{equation}
%\pb{\pi^{ij}(\bx),g^{kl}(\by)}=
%%  -g^{km}
%% (\by)\pb{\pi^{ij}(\bx),g_{mn}(\by)}
%% g^{nl}(\by)
%\frac{1}{2}( g^{ik}g^{jl}+
%g^{jk}g^{il})\delta(\bx-\by)
%\end{equation}
%we find that the Poisson bracket
%of $Q^{\dag ij}$ with $Q^{kl}$ is nonzero
\begin{eqnarray}
\pb{Q^{\dag ij}(\bx),Q^{kl}(\by)}=
% -i\frac{\Lambda}{4}
% (g^{ik}g^{jl}+g^{il}g^{kl})\sqrt{g}\delta
% (\bx-\by)-
% \nonumber \\
% -i\frac{\Lambda}{8}g^{kl}g^{ij}
% \sqrt{g}(\bx)\delta(\bx-\by)+
% i\frac{\Lambda}{8}g^{kl}g^{ij}
% \delta(\bx-\by)=\nonumber \\
-i\frac{\Lambda}{4}
(g^{ik}g^{jl}+g^{il}g^{kl})\sqrt{g}\delta
(\bx-\by) \ .  \nonumber \\\
\end{eqnarray}
It is important that this Poisson
bracket  is proportional to
$\delta(\bx-\by)$ and does not contain
 derivative of delta function.
% Further, we again find
%that
%\begin{eqnarray}
%\pb{Q^{\dag ij}(\bx),\frac{1}{g}
%\mG_{klmn}(\by)}=
%-i\pb{\pi^{ij}(\bx),\frac{1}{g}
%\mG_{klmn}(\by)}=
%-\pb{Q^{ij}(\bx),
%\frac{1}{g}\mG_{klmn}(\by)} \ .
%\nonumber \\
%\end{eqnarray}
Then with the help of this result and
the second equation in (\ref{QQul}) we
again find that
\begin{eqnarray}
\pb{Q^{\dag ij}\frac{1}{g}
\mG_{ijkl}Q^{kl}(\bx),
Q^{\dag ij}\frac{1}{g}
\mG_{ijkl}Q^{kl}(\by)}
% \pb{Q^{\dag ij}(\bx),\frac{1}{g}
% \mG_{ijkl}(\by)}Q^{kl}(\bx)
% \frac{1}{g}\mG_{ijkl}(\by)Q^{\dag ij}(\by)
% Q^{kl}(\by)+\nonumber \\
% +Q^{\dag ij}(\bx)Q^{kl}(\bx)
% \pb{\frac{1}{g}\mG_{ijkl}(\bx),Q^{\dag ij}(\by)}
% \frac{1}{g}\mG_{ijkl}(\by)Q^{kl}(\by)+\dots
% \nonumber \\
% =\pb{Q^{\dag ij}(\bx),\frac{1}{g}
% \mG_{ijkl}(\by)}Q^{kl}(\bx)
% \frac{1}{g}\mG_{ijkl}(\by)Q^{\dag ij}(\by)
% Q^{kl}(\by)-\nonumber \\
% -Q^{\dag ij}(\bx)Q^{kl}(\bx)
% \pb{Q^{\dag ij}(\by),\frac{1}{g}\mG_{ijkl}(\bx)}
% \frac{1}{g}\mG_{ijkl}(\by)Q^{kl}(\by)+\dots
=0 \   \nonumber \\
\end{eqnarray}
and as follows from (\ref{mH0ul}) the
Poisson brackets of the Hamiltonian
constraints vanish.

In summary, the ultralocal $f(R)$
Ho\v{r}ava-Lifshitz gravity has the
same nice property as the ultralocal
theory of gravity \cite{Isham:1975ur}.

{\bf Acknowledgements:} This work was
 supported by the Czech
Ministry of Education under Contract
No. MSM 0021622409.

\newpage
%%%%%%%%%%%%%%%%%%%%%%%%%%%%%%%%%%%%%%
%%%%%%% Thebibligraphy %%%%%%%%%%%%%%%%%%%%%
%%%%%%%%%%%%%%%%%%%%%%%%%%%%%%%%%%%%%

\end{document}